\documentstyle[aps,prl,multicol]{revtex}

\newcommand{\ns}[1]{\hbox{$\!\!\!#1\!\!\!$}}

\def\inn{{\text{in}}} \def\out{{\text{out}}}

\def\lb{{\hbox{[}}} \def\rb{{\hbox{] }}}

\def\d{\displaystyle }
\begin{document}

\title{%
  Another Note on Forced Burgers Turbulence} \author{Weinan E$^{a)}$
  and Eric Vanden Eijnden$^{b)}$}
\address{Courant Institute of Mathematical Sciences\\
  New York University\\
  New York, New York 10012}

\maketitle

\begin{abstract}
  The power law range for the velocity gradient probability density
  function in forced Burgers turbulence has been an issue of intense
  discussion recently.  It is shown in \lb chao-dyn/9901006\rb that
  the negative exponent in the assumed power law range has to be
  strictly larger than $3$.  Here we give another direct argument for
  that result, working with finite viscosity.  At the same time we
  discuss viscous corrections to the power law range.  This should
  answer the questions raised by Kraichnan in \lb chao-dyn/9901023\rb
  regarding the results of \lb chao-dyn/9901006\rb.
\end{abstract}

\vskip .2cm

\begin{multicols}{2}

  The main purpose of this note is to clarify and extend to finite
  viscosity the results of our earlier paper \cite{eva99} concerning
  the asymptotic behavior of the velocity gradient probability density
  function (PDF) for Burgers turbulence with homogeneous, smooth,
  Gaussian, and white-in-time forcing. In particular, the present note
  should answer the questions raised in \cite{kra99} regarding
  \cite{eva99}.
  
  Let $Q^\nu(\xi,t)$ denote the PDF of $\xi=u_x$, where $u$ satisfies
\begin{equation}
  \label{eq:1}
  u_t+u u_x = \nu u_{xx} + f,
\end{equation}
and define
\begin{equation}
  \label{eq:2}
  Q(\xi,t)=\lim_{\nu\to0} Q^\nu(\xi,t).
\end{equation}
Then the question of interest is the value of $\alpha$, such that
\begin{equation}
  \label{eq:3}
  Q\sim C_-|\xi|^{-\alpha} \qquad 
  \text{as} \ \ \xi\to-\infty.
\end{equation}
We emphasize that the existence of such a range (the so-called
$-\alpha$ range) is not the issue here.  The issue is the value of
$\alpha$.  Notice that (\ref{eq:3}) is a statement about the inviscid
limit.  For fixed $\nu>0$, the left tail of $Q^\nu$ decays much faster
due to the presence of the viscous range. In addition $Q^\nu$
satisfies
\begin{equation}
  \label{eq:4}
    Q^\nu_t=\xi Q^\nu +\bigl(\xi^2 Q^\nu\bigr)_\xi +B_1 Q^\nu_{\xi\xi}
      +F^\nu,
\end{equation}
where $F^\nu = -\nu\bigl\langle
\xi_{xx}\delta[\xi-\xi(x,t)]\bigr\rangle_\xi$ accounts for the effect
of the viscosity, and $B_1=\int_0^{\infty} dt\, \langle
f_x(x,t)f_x(x,0)\rangle$.  $Q$~satisfies an equation similar to
(\ref{eq:4}) with $F^\nu$ replaced by $F=\lim_{\nu\to0} F^\nu$:
\begin{equation}
  \label{eq:5}
    Q_t=\xi Q +\bigl(\xi^2 Q\bigr)_\xi + B_1 Q_{\xi\xi} +F.
\end{equation}
The expression for $F$ is given explicitly in (\ref{eq:12}).

One main result of \cite{eva99} is a statement to the effect that
$\alpha>3$, expressed as
\begin{equation}
  \label{eq:6}
  \lim_{\xi\to-\infty} \xi^3 Q(\xi,t)=0.
\end{equation}
We emphasize that there is an important distinction between the strict
inequality $\alpha>3$ and the bound $\alpha\ge3$ advanced in
\cite{kra99}: (\ref{eq:6}) rules out all the predictions in the
literature (including those of \cite{pol95,bol97,gokr98}) except that
of \cite{ekhma97} with $\alpha=7/2$.  As is discussed in \cite{kra99},
$\alpha = 3$ and $\alpha > 3$ imply qualitatively different picture
concerning the contribution of the $F^{\nu}$ term in (\ref{eq:4}) in
the viscous range.

In \cite{eva99}, (\ref{eq:6}) was derived from (\ref{eq:5}). Here we
will work directly with (\ref{eq:4}).  Consider statistical steady
state ($Q^\nu_t=0$). For $\xi\ll-\xi_0$ ($=-B_1^{1/3})$, the $B_1$
term in (\ref{eq:4}) can be neglected giving
\begin{equation}
  \label{eq:7}
  \xi Q^\nu +\bigl(\xi^2 Q^\nu\bigr)_\xi +F^\nu\approx 0,
\end{equation}
or, equivalently,
\begin{equation}
  \label{eq:8}
  \bigl(\xi^3 Q^\nu\bigr)_\xi \approx -\xi F^\nu.
\end{equation}
Because of the exponential decay in the viscous range, we get from
(\ref{eq:8})
\begin{equation}
  \label{eq:9}
  Q^\nu 
  \sim |\xi|^{-3} \int_{-\infty}^\xi \!\! d\xi'\ \xi' F^\nu(\xi')\qquad
  \text{for} \ \ \xi\ll-\xi_0.
\end{equation}
Using the same analysis as in \cite{eva99}, for small $\nu$ we can
obtain an explicit expression for $Q^\nu$ from (\ref{eq:9}).  The
calculation is performed in Appendix~\ref{sec:A} and gives
\begin{displaymath}
  \begin{array}{rcl}
    {\d Q^\nu}&\ns{\sim}&{\d
      \rho|\xi|^{-3}  \int_\xi^{+\infty} \!\!\! d\xi'\!
      \int_{-\infty}^{s_\star}\!\!
      ds\ 
      \xi' \sqrt{\case{1}{4}s^2+2\nu(\xi-\xi')}\, V(\xi',s)}\\[8pt]
    &\ns{-}&{\d\nu \rho|\xi|^{-2}  \int_\xi^{+\infty}\! \!\! d\xi'\! 
      \int_{-\infty}^{s_\star}\!\!\!
      ds \frac{(\xi-\xi')}
      {\sqrt{\case{1}{4}s^2+2\nu(\xi-\xi')}}V(\xi',s)}
  \end{array}
\end{displaymath}
\vspace{-15truept}
\begin{equation}
  \label{eq:10}
   \qquad\qquad\qquad\qquad\qquad  \text{for} \ \ \xi\ll-\xi_0,
\end{equation}
where $\rho$ is the number density of shocks,
$s_\star=-2\sqrt{2\nu(\xi'-\xi)}$, $V(\xi,s)=V_-(\xi,s)+V_+(\xi,s)$,
$V_\pm(\xi_\pm,s)$ are the PDFs of $\xi_\pm(y,t)$ (gradients at the
left and right of the shock in the inviscid limit) and $s(y,t)$ (shock
amplitude, $s\le0$), conditional on the property that there is a shock
at $x=y$.

Without further information about $V(\xi,s)$ it is difficult to carry
out asymptotics on (\ref{eq:10}), and we shall not dwell on this
problem (see however (\ref{eq:A15})-(\ref{eq:A18}) in
Appendix~\ref{sec:A}). What is easier and more instructive is to
actually take the limit as $\nu\to0$ for fixed $\xi$ in (\ref{eq:10}).
Then, only the first term at the right hand-side of (\ref{eq:10})
survives, and $Q^\nu$ converges to
\begin{equation}
  \label{eq:11}
  Q(\xi)\sim \xi^{-3} \int_\xi^{+\infty} \!\! d\xi'\ \xi' F(\xi') \qquad
  \text{for} \ \ \xi\ll-\xi_0,
\end{equation}
with
\begin{equation}
  \label{eq:12}
  F(\xi)=\lim_{\nu\to0} F^\nu(\xi)=\frac{\rho}{2}
      \int_{-\infty}^0\!\! ds\ s\, V(\xi,s).
\end{equation}
At this stage, we use the fact that
\begin{equation}
  \label{eq:13}
  A=\int_{-\infty}^{+\infty}\!\!d\xi \ \xi F =
  \frac{\rho}{2}[ \langle s\xi_-\rangle+\langle s\xi_+\rangle]=0.
\end{equation} 
This is a steady state consequence of
\begin{equation}
  \label{eq:14}
  \frac{d}{dt}(\rho\langle s\rangle)=-
  \frac{\rho}{2}[ \langle s\xi_-\rangle+\langle s\xi_+\rangle],
\end{equation}
which is proven in Appendix~\ref{sec:B}. As a result of (\ref{eq:13}),
(\ref{eq:11}) may be rewriten as
\begin{equation}
  \label{eq:15}
  Q(\xi)\sim |\xi|^{-3} \int_{-\infty}^\xi \!\! d\xi'\ \xi' F(\xi') \qquad
  \text{for} \ \ \xi\ll-\xi_0.
\end{equation}
(\ref{eq:6}) then follows because $\int_{-\infty}^{\xi} d\xi' \xi'
F(\xi') \to 0$ as $\xi \to -\infty$.  We stress that this argument,
and in particular the use of (\ref{eq:14}) at steady state, does not
imply that $Q\sim A|\xi|^{-3}$ at transient states because the whole
analysis here rests on (\ref{eq:9}) which is only valid at statistical
steady state. In fact, as we will now show, (\ref{eq:6}) also holds
for transient states\cite{rem1}. Again in \cite{eva99} this argument
was presented for $Q$.  Here we reinterpret this using $Q^\nu$ for
small $\nu$.

Multiplying (\ref{eq:4}) by $\xi$ and integrating from $\xi_\star$ to
$+\infty$ (for fixed $\xi_\star\ll-\xi_0$), we get
\begin{equation}
  \label{eq:16}
  \int_{\xi_\star}^{+\infty}\!\!d\xi \ \xi Q^\nu_t 
  = -\xi_\star^3 Q^\nu(\xi_\star,t) 
  +\int_{\xi_\star} ^{+\infty}\!\! d\xi \ \xi F^\nu.
\end{equation}
Here we neglected contribution from the $B_1$ term since it is small
compared with the remaining terms.  Taking the limit as $\nu \to 0$,
we get:
\begin{equation}
  \label{eq:17}
  \int_{\xi_\star}^{+\infty}\!\!d\xi \ \xi Q_t = -\xi_\star^3 Q(\xi_\star,t)
  +\int_{\xi_\star} ^{+\infty}\!\! d\xi \ \xi F.
\end{equation}
Therefore
\begin{equation}
  \label{eq:18}
  \lim_{\xi_\star \to -\infty} \xi_\star^3 Q(\xi_\star,t) 
  = \int_{-\infty}^{+\infty}\!\!d\xi\
  \xi F - \frac{d}{dt}\int_{-\infty}^{+\infty}\!\!d\xi\ \xi Q.
\end{equation}
Notice that even though by homogeneity
\begin{equation}
  \label{eq:19}
  \langle \xi\rangle_\nu= 
      \int_{-\infty}^{+\infty}\!\!d\xi \ \xi Q^\nu=0,
\end{equation}
in the limit as $\nu\to0$
\begin{equation}
  \label{eq:20}
  \langle \xi\rangle = \int_{-\infty}^{+\infty}\!\!d\xi \ \xi Q
  =-\rho \langle s \rangle \neq 0.
\end{equation}
In other words, a finite amount of $\xi = u_x$ has gone to the shocks
in the limit as $\nu \to 0$.  Hence, using (\ref{eq:20}) at the right
hand side of (\ref{eq:18}), this equation becomes
\begin{equation}
  \label{eq:18b}
  \begin{array}{rcl}  
  {\d\lim_{\xi_\star \to -\infty} \xi_\star^3 Q(\xi_\star,t)}&\ns{=}& 
  {\d\frac{\rho}{2}[ \langle s\xi_-\rangle+\langle s\xi_+\rangle]-
  \frac{d}{dt}\langle\xi\rangle}\\[4pt]
  &\ns{=}& 
  {\d\frac{\rho}{2}[ \langle s\xi_-\rangle+\langle s\xi_+\rangle]+
  \frac{d}{dt}(\rho\langle s\rangle)}\\[6pt]
  &\ns{=}& 
  {\d0.}
  \end{array}
\end{equation}
The last equality follows from (\ref{eq:14}).

One main message in \cite{kra99} is the claim that the argument in
\cite{eva99} which led to the strict bound (\ref{eq:6}) is
insufficient.  Here we paraphrase the argument which led to this
claim.  One may always write (\ref{eq:9}) as
\begin{equation}
  \label{eq:21}
  |\xi|^{3}Q^\nu(\xi)=\int_{-\infty}^{\xi_M} \!\! d\xi'\ \xi' F^\nu(\xi')
  +\int_{\xi_M}^\xi \!\! d\xi'\ \xi' F^\nu(\xi'),
\end{equation}
where $\xi_M$ is defined as the value at which the $-\alpha$ range is
masked by the viscous range. Assuming the latter behaves as $\nu
C|\xi|^{-1}$ (see \cite{kra99} and (\ref{eq:A16})) then $\xi_M$ is
determined from solving
\begin{equation}
  \label{eq:22}
  C_-|\xi|^{-\alpha} = \nu C|\xi|^{-1},
\end{equation}
which gives $\xi_M = -C_0\nu^{-1/(\alpha-1)}$ with $C_0=
(C/C_-)^{-1/(\alpha-1)}$.  In the limit as $\nu\to0$, the second term
at the right-hand of (\ref{eq:21}) gives
\begin{equation}
  \label{eq:23}
  \lim_{\nu\to0}\int_{\xi_M}^{\xi} \!\! d\xi'\ \xi' F^\nu(\xi')=
  \int_{-\infty}^{\xi} \!\! d\xi'\ \xi' F(\xi'),
\end{equation}
which goes to $0$ as $\xi\to-\infty$.  Therefore, whether $\alpha>3$
depends on whether
\begin{equation}
  \label{eq:24}
  \lim_{\nu\to0} \int_{-\infty}^{\xi_M} \!\! d\xi'\ \xi' F^\nu(\xi')=0,
\end{equation}
holds. Since
\begin{equation}
  \label{eq:25}
  \xi_M^3 Q^\nu(\xi_M)=\int_{-\infty}^{\xi_M} \!\! d\xi'\ \xi' F(\xi'),
\end{equation}
an equivalent form of (\ref{eq:24}) is
\begin{equation}
  \label{eq:26}
  \lim_{\nu\to0} \xi_M^3 Q^\nu(\xi_M)=0.
\end{equation}
Although this statement is correct and gives another way of
appreciating the difference between $\alpha = 3$ and $\alpha > 3$, it
is an unproductive approach of addressing the issue of whether $\alpha
=3$, for the simple reason that the validity of (\ref{eq:24}) and
(\ref{eq:26}) depends sensitively on the value of $\xi_M$ which cannot
be known prior to knowing $\alpha$.  Unlike what is claimed in
\cite{kra99}, $\xi_M$ cannot be an arbitrary choice that satisfies
$\xi_0/\xi_M\to0$, $\nu\xi_M\to0$ as $\nu\to0$. For example if we
choose $\xi_M=\xi_N=-C_1 \nu^{-1/2} $, from (\ref{eq:A16})
\begin{equation}
  \label{eq:27}
  \lim_{\nu\to0} \xi^3_N Q^\nu (\xi_N) = C C_1^2 \not=0,
\end{equation}
regardless the value of $\alpha$.  On the other hand, it is easy to
see that if $\xi_M = o(\nu^{-1/2})$, then (\ref{eq:24}) and
(\ref{eq:26}) hold.  The important technical point in \cite{eva99} is
to find ways to circumvent this path. That was done by studying
directly the inviscid limit of $Q^\nu$. Here we have presented a
direct argument based on the integral expression for $Q^\nu$.  As a
by-product, we have
\begin{equation}
  \label{eq:28}
  \xi_M = o(\nu^{-1/2}).
\end{equation}

It is also worth stressing that even though (\ref{eq:9}) and
(\ref{eq:15}) look similar, they are not equivalent.  (\ref{eq:9}) is
meant for the case of finite $\nu$ and is derived using
straightforward integration from (\ref{eq:7}).  (\ref{eq:15}) is valid
for the limiting PDF $Q$ and its derivation is much more non-trivial.
In Appendix~\ref{sec:C}, we show that (\ref{eq:15}) can be derived
directly from (\ref{eq:5}) using the realizability constraint over the
{\em whole $\xi$-line}, as well as the additional information provided
by (\ref{eq:14}). It shows that the only realizable steady state
solutions of (\ref{eq:5}) has the form (\ref{eq:15}). Other solutions
violate non-negativity either for $\xi\to-\infty$ or for
$\xi\to+\infty$.  The argument in \cite{eva99} was a global argument,
not localized at very large negative values of $\xi$.

The characterization that ``the analysis in \cite{eva99} is done in
terms of a split of $u$ and $\xi$ into a part exterior to shocks and a
part interior to shocks''\cite{kra99} also needs more clarification.
What was done in \cite{eva99} was a derivation of an approximation to
$\nu\xi_{xx}$ (or $\nu\xi_x^2$) using boundary layer analysis and
matched asymptotics in order to evaluate the limit of $F^\nu$ as
$\nu\to0$.  The same technique was used here to evaluate directly the
limit of $Q^\nu$ (for large negative $\xi$) as $\nu\to0$.  This
approximation is uniformly valid except at shock creation and
collision whose contributions to $F^\nu$ is of lower order. It can
also be systematically improved if additional information is required.

Going back to statistical stationary state, what is the actual value
of $\alpha$?  \cite{ekhma97} predicted that $\alpha = 7/2 $ under the
assumption that the main contribution to $Q$ for large negative values
of $\xi$ comes from neighborhoods of shock creation points
(pre-shocks).  This geometric argument was expanded in \cite{eva99} in
the context of (\ref{eq:5}) and in particular the form of $F$: under
the geometric argument $ F(\xi) \sim C |\xi|^{-5/2}$ with $C<0$ for
$\xi \ll -\xi_0$.  \cite{kra99} further expanded the geometric
argument and obtained values of $\alpha$ in $(3, 7/2)$ by considering
special singular data.  Polyakov \cite{pol98} gave an example that
gives $\alpha =3$.  However these are rather pathological situations
that lie outside the regime of interest here, i.e. the case of smooth
forcing.  By studying the master equation for the environment of
shocks, \cite{eva299} verifies that indeed the main contribution does
come from shock creation points, and thereby confirms $\alpha = 7/2$.

In conclusion, we stress that there are many ways to exclude the
possibility of having $|\xi|^{-3}$ behavior for the left tail of $Q$
in the case of smooth Gaussian force. The discussion in \cite{kra99}
provides yet another way of understanding the different consequences
of $\alpha=3$ and $\alpha>3$, but it is not the right way to address
the issue of whether $\alpha=3$.

\section*{Acknowledgments}

We benefited from our frequent e-mail exchanges with R. H. Kraichnan
and A. M. Polyakov. The work of E is supported by a Presidential
Faculty Fellowship from the National Science Foundation.  The work of
Vanden Eijnden is supported by U.S. Department of Energy Grant No.
DE-FG02-86ER-53223.

\appendix

\section{Evaluation of $Q^\nu$}
\label{sec:A}

In this appendix we perform the computation of $Q^\nu$ to $O(\nu)$.
We first put (\ref{eq:9}) in a form which is more convenient for the
calculation. Notice that
\begin{equation}
  \label{eq:A1}
  F^\nu(\xi,t)=G^\nu_{\xi\xi}(\xi,t),
\end{equation}
where
\begin{equation}
  \label{eq:A2}
  G^\nu(\xi,t)= -\nu\bigl\langle\xi_{x}^2(x,t)
      \delta[\xi-\xi(x,t)]\bigr\rangle.
\end{equation}
Thus, from (\ref{eq:9}),
\begin{equation}
  \label{eq:A3}
  Q^\nu 
  \sim -|\xi|^{-3} G^\nu-|\xi|^{-2} G^\nu_\xi \qquad
  \text{for} \ \ \xi\ll-\xi_0.
\end{equation}
We now evaluate $G^\nu$.  For statistically homogeneous situations,
the averages at the right hand-side of (\ref{eq:A2}) can be evaluated
upon resorting to spatial ergodicity to replace the ensemble-average
by the space-average:
\begin{equation}
  \label{eq:A4}
  G^\nu(\xi,t) =-\nu\lim_{L\to\infty}\frac{1}{2L} \int_{-L}^{L} dx\ \xi_{x}^2 
      \delta[\xi-\xi(x,t)].
\end{equation}
In the limit of small $\nu$ only small intervals around the shocks
will contribute to this integral. In these layers, we use boundary
layer analysis to evaluate $\xi(x,t)$. This analysis was outlined in
\cite{eva99} (for details see \cite{eva299}): Let $y$ be a shock
position. Near $y$, $u$ can be expressed as
\begin{equation}
  \label{eq:A5}
  u(x,t)=u^\inn(x,t)=v\left(\frac{x-y}{\nu},t\right),
\end{equation}
and the expression for $v(z,t)$ can be obtained via a series expansion
in $\nu$. It yields $v=v_0+\nu v_1+O(\nu^2)$, with
\begin{equation}
  \label{eq:A6}
  v_0(z,t)= \bar u-\frac{s}{2} \tanh \left(\frac{sz}{4}\right).
\end{equation}
$v_1$ is a solution of
\begin{equation}
  \label{eq:A7}
  {v_0}_t+(v_0-\bar u){v_1}_z+v_1 {v_0}_z ={v_1}_{zz}+f.
\end{equation}
The actual expression of $v_1$ is rather complicated, and the only
information really needed about $v_1$ to evaluate (\ref{eq:A4}) is the
values of ${v_1}_z$ as $z\to\pm\infty$. Let $\lim_{z\to\pm\infty}
{v_1}_z= \xi_\pm$. Then
\begin{equation}
  \label{eq:A8}
  \xi_\pm = \mp \frac{2\bar u_t}{s}-\frac{s_t}{s} 
  \pm\frac{2f}{s},
\end{equation}
or, equivalently,
\begin{equation}
  \label{eq:A9}
  s_t=-\frac{s}{2}  \bigl( \xi_- +\xi_+\bigr),\qquad
  \bar u_t=\frac{s}{4}  \bigl( \xi_- -\xi_+\bigr)+f.
\end{equation}
In these expressions, the values of $s$ and $\bar u=dy/dt$ must be
obtained by matching $u^\inn(x,t)$ with the solution of the Burgers
equation outside the shock layer, say, $u^\out(x,t)$: this eventually
produces an approximation for $u(x,t)$ uniformly valid except at shock
creation and collision.

Using the results of the boundary layer analysis, to $O(\nu)$
(\ref{eq:A4}) can be estimated as ($\xi^\inn=u^\inn_x$)
\begin{equation}
  \label{eq:A10}
  \begin{array}{l}
    {\d G^\nu(\xi,t) }\\[4pt]
    {\d=-\nu\lim_{L\to\infty}\frac{N}{2L} \frac{1}{N}
      \sum_{j} \int_{\rm j^{th} layer}\hskip-5mm  dx\
      (\xi^\inn_{x})^2\, \delta[\xi-\xi^\inn(x,t)],}
  \end{array}
\end{equation}
or, picking any particular shock layer, going to the stretched
variable $z=(x-y)/\nu$, and taking the limit as $L\to\infty$,
\begin{equation}
  \label{eq:A11}
  \begin{array}{rcl}
    {\d G^\nu(\xi,t)}&\ns{=}&
    {\d-\rho
      \int ds d\bar u d\xi_-d\xi_+\
      T(\bar u,s,\xi_-,\xi_+,s,t)}\\[4pt]
    &&{\d\qquad\times \int_{-\infty}^{+\infty}\!\! dz
      \ \eta^2_{z}\, 
      \delta[\xi-\eta(z,t)],}
  \end{array}
\end{equation}
where we defined $\eta= \nu^{-1}{v_0}_{z}+{v_1}_{z}\approx
\nu^{-1}{v_0}_{z}+\xi_\pm$. Here $T(\bar u,s,\xi_-,\xi_+,t)$ is the
PDF of $\bar u$, $s$, $\xi_-$, $\xi_+$ conditional on the existence of
a shock at $x=y$; it arises because $\eta(z,t)$ depends parametrically
on these (random) variables. To perform the $z$ integral in
(\ref{eq:A11}), we change the integration variable to
$\xi'=\eta-\xi_-$ for $z<0$ and to $\xi'=\eta-\xi_+$ for $z>0$.  Then,
using the equation for $v_0$, $(v_0-\bar u){v_0}_z={v_0}_{zz}$, as
well as (\ref{eq:A6}), we have $\eta_{z}=[(v_0-\bar u)(\eta-\xi_\mp)]=
\pm[\sqrt{s^2/4+2\nu(\eta-\xi_\mp)}(\eta-\xi_\mp)]$, and
\begin{equation}
  \label{eq:A12}
  dz \ \eta^2_{z}=\pm d\xi' \ \xi'\sqrt{\case{1}{4}s^2+2\nu\xi'}.
\end{equation}
Thus
\begin{equation}
  \label{eq:A13}
  \begin{array}{l}
  {\d -\int_{-\infty}^{+\infty}\!\! dz
    \ \eta^2_{z}\, \delta[\xi-\eta(z,t)]}\\[6pt]
  {\d=\int_{-s^2/8\nu}^0\!\!\!d\xi'
      \ \xi'\sqrt{\case{1}{4}s^2+2\nu\xi'}
      \ \delta[\xi-\xi'-\xi_-]}\\[6pt]
  {\d+\int_{-s^2/8\nu}^0\!\!\!d\xi'
      \ \xi'\sqrt{\case{1}{4}s^2+2\nu\xi'}
      \ \delta[\xi-\xi'-\xi_+]}\\[10pt]
  {\d=(\xi-\xi_-)\sqrt{\case{1}{4}s^2+2\nu(\xi-\xi_-)}}\\[6pt]
  {\d \qquad  \times 
    \bigl[\theta(\xi_--\xi)-\theta(\xi_--s^2/8\nu-\xi)\bigr]}\\[6pt]
  {\d+(\xi-\xi_+)\sqrt{\case{1}{4}s^2+2\nu(\xi-\xi_+)}}\\[6pt]
  {\d \qquad  \times 
    \bigl[\theta(\xi_+-\xi)-\theta(\xi_+-s^2/8\nu-\xi)\bigr],}
    \end{array}
\end{equation}
where $\theta(x)$ is the Heaviside function.  Inserting this
expression in (\ref{eq:A11}) then results in
\begin{equation}
  \label{eq:A14}
  \begin{array}{l}
    {\d G^\nu(\xi,t)= }\\[4pt]
    {\d\rho
      \int_\xi^{+\infty}\!\! d\xi' \int_{-\infty}^{s_\star}\!\!\! ds \
      (\xi-\xi')\sqrt{\case{1}{4}s^2+2\nu(\xi-\xi')}V(\xi',s,t),}\\[4pt]
  \end{array}
\end{equation} 
where we used $\int d\bar u d\xi_\mp\, T(\bar u,s,\xi_-,\xi_+,t)=
V_\pm(\xi_\pm,s,t)$ and $s_\star=-2\sqrt{2\nu(\xi'-\xi)}$. Inserting
(\ref{eq:A14}) in (\ref{eq:A3}) gives (\ref{eq:10}).

Note that if one neglects the $O(1)$ term in the expansion for
$\xi^\inn$ (an assumption we do {\em not} make), then
$V(\xi,s,t)=2S(s,t)\delta(\xi)$, where $S(s,t)$ is the conditional PDF
of $s(y,t)$. At statistical steady state this is
$V(\xi,s)=2S(s)\delta(\xi)$ and (\ref{eq:10}) reduces to
\begin{equation}
  \label{eq:A15}
  Q^\nu\sim 2\nu \rho|\xi|^{-1}  
      \int_{-\infty}^{-2\sqrt{2\nu|\xi|}}\!\!\!
      ds \frac{S(s)}
      {\sqrt{\case{1}{4}s^2+2\nu\xi}}.
\end{equation}
This is the expression obtained by Gotoh and Kraichnan \cite{gokr98}
for $Q^\nu$ in the viscous range. (\ref{eq:A15}) is hard to justify
since it amounts to assessing the accuracy of the approximation
$V(\xi,s,t)\approx 2S(s,t)\delta(\xi)$ for large negative $\xi$.
Granting (\ref{eq:A15}), $Q^\nu$ can be further simplified if one
assumes that the first inverse moment of the shock amplitude $s$ is
finite, i.e.  $\langle|s|^{-1}\rangle=-\int_{-\infty}^0ds
S(s)/s<+\infty$.  Then, for $\xi=o(\nu^{-1})$, (\ref{eq:A15}) reduces
to
\begin{equation}
  \label{eq:A16}
    Q^\nu \sim \nu C |\xi|^{-1},
\end{equation}
\begin{equation}
  \label{eq:A17}
  C = 4\rho \bigl\langle |s|^{-1}\bigr\rangle.
\end{equation}
(\ref{eq:A16}) describes the well-known $|\xi|^{-1}$ viscous range
(the $-1$~range). This expression can be combined with (\ref{eq:15})
to give
\begin{equation}
  \label{eq:A18}
   Q^\nu \sim C_-|\xi|^{-\alpha}+\nu C |\xi|^{-1},
\end{equation}
on the range $\xi=o(\nu^{-1})$, $\xi\ll-\xi_0$. Here
$C_-|\xi|^{-\alpha}=|\xi|^{-3}\int_{-\infty}^\xi\!\!d\xi'\ \xi'
F(\xi')$.

\section{Derivation of (14)}
\label{sec:B}

To get (\ref{eq:14}) we need to evaluate the time derivative of (using
ergodicity)
\begin{equation}
  \label{eq:B1}
  \rho\langle s\rangle = \lim_{L\to\infty}\frac{N}{2L} \frac{1}{N} 
      \sum_{j=1}^N s(y_j,t),
\end{equation}
where $N$ is the number of shocks in $[-L,L]$ and the $y_j$'s are
their locations.  Using (\ref{eq:A9}), clearly the time derivative of
(\ref{eq:B1}) will give (\ref{eq:14}) if the time dependence of $N$
does not make any contribution. $N$ varies due to shock creation or
shock collision. Consider the creation first, and assume a shock is
created at position $y_1$ at time $t_1$. Then one has in (\ref{eq:B1})
a term like (disregarding the factor $1/2L$)
\begin{equation}
  \label{eq:B2}
  T_1=s(y_1,t) \theta(t-t_1),
\end{equation}
where $\theta(x)$ is the Heaviside function.  Time differentiation of
(\ref{eq:B2}) gives
\begin{equation}
  \label{eq:B3}
  \frac{dT_1}{dt}=\frac{d}{dt}s(y_1,t) \theta(t-t_1)+ 
  s(y_1,t) \delta(t-t_1).
\end{equation}
The second term accounts for the time dependence of $N$.  Since the
shock amplitude is zero at creation, $s(y_1,t)
\delta(t-t_1)=s(y_1,t_1) \delta(t-t_1)=0$.  This means that the term
$dN/dt$ makes no contribution to the time derivative of (\ref{eq:B1})
at shock creation.  Consider now the merging events. Assume the shocks
located at position $y_2$ and $y_3$ merge into one shock located at
position $y_1$ at time $t_1$. By definition
$y_1(t_1)=y_2(t_1)=y_3(t_1)$.  Such an event contributes in
(\ref{eq:B1}) by a term like
\begin{equation}
  \label{eq:B4}
  T_2=s(y_1,t) \theta(t-t_1)+\bigl[s(y_2,t)+s(y_3,t)\bigr]\theta(t_1-t).
\end{equation}
Time differentiation of (\ref{eq:B4}) gives
\begin{equation}
  \label{eq:B5}
  \begin{array}{rcl}
    {\d\frac{dT_2}{dt}}&\ns{=}&{\d
      \frac{d}{dt}s(y_1,t) \theta(t-t_1)}\\[6pt]
    &\ns{+}&{\d \frac{d}{dt}
      \bigl[s(y_2,t)+s(y_3,t)\bigr]\theta(t_1-t)}\\[8pt]
    &\ns{+}&{\d s(y_1,t) \delta(t-t_1)-\bigl[s(y_2,t)+s(y_3,t)\bigr]
      \delta(t-t_1).}
  \end{array}
\end{equation}
Since shock amplitudes add up at collision:
\begin{equation}
  \label{eq:B6}
  \begin{array}{cl}
    &{\d\lim_{t\to0+}s(y_1(t_1+t),t_1+t)}\\[6pt]
    \ns{=}&{\d\lim_{t\to0+}
      \bigl[s(y_2(t_1-t),t_1-t)+s(y_3(t_1-t),t_1-t)\bigr].}
  \end{array}
\end{equation}
Thus the terms in (\ref{eq:B5}) involving $\delta$-functions vanish.
This means that the term $dN/dt$ makes no contribution to the time
derivative of (\ref{eq:B1}) at shock collision. Hence (\ref{eq:14}).

\section{Global realizability constraints}
\label{sec:C}

Here we study (\ref{eq:5}) at steady state
\begin{equation}
  \label{eq:C1}
  0=\xi Q +\bigl(\xi^2 Q\bigr)_\xi+
  B_1 Q_{\xi\xi}+F.
\end{equation} 
We will show that the only non-negative solution of (\ref{eq:C1}) is
\begin{equation}
  \label{eq:C2}
  Q_s(\xi)= \frac{1}{B_1}\int_{-\infty}^\xi \!\! d\xi'
      \ \xi' F(\xi')-\frac{\xi{\rm e}^{-\Lambda}}{B_1}
      \int_{-\infty}^\xi \!\! d\xi'
      \ {\rm e}^{\Lambda'} G(\xi'),
\end{equation}
where $\Lambda=\xi^3/3B_1$ and
\begin{equation}
  \label{eq:C3}
  G(\xi)=F(\xi)+\frac{\xi}{B_1} \int_{-\infty}^{\xi}d\xi'\ \xi'
      F(\xi').
\end{equation}
We will also show that for large positive $\xi$, $F$ decays in such a
way that
\begin{equation}
  \label{eq:C4}
  \lim_{\xi\to+\infty} \xi^{-2} {\rm e}^{\Lambda} F(\xi)=0,
\end{equation}
assuming the limit exists.  Note that from (\ref{eq:13}), the first
moment of $F$ exists. Hence $\xi F(\xi)$ is integrable at $-\infty$.

Before proving (\ref{eq:C2}) and (\ref{eq:C4}), let us consider some
simple consequences.  For $\xi<0$, integrating by parts in
(\ref{eq:C2}) gives
\begin{equation}
  \label{eq:C5}
  Q_s(\xi)=
  -\frac{\xi{\rm e}^{-\Lambda}}{B_1^2}\int_{-\infty}^\xi \!\! d\xi'
  \ \frac{{\rm e}^{\Lambda'}}{\xi'^2}
  \int_{-\infty}^{\xi'}d\xi'' \ \xi'' F(\xi'').
\end{equation}
Similarly for $\xi>0$, we have
\begin{equation}
  \label{eq:C6}
  Q_s(\xi)=C_+\xi{\rm e}^{-\Lambda}
  -\frac{\xi{\rm e}^{-\Lambda}}{B_1}\int_\xi^{+\infty} \!\!\! d\xi'
  \ \frac{{\rm e}^{\Lambda'}}{\xi'^2}
  \int_{\xi'}^{+\infty}\!\!\! d\xi'' \ \xi'' F(\xi''),
\end{equation}
with
\begin{equation}
  \label{eq:C7}
  C_+=-\frac{1}{B_1}\int_{-\infty}^{+\infty}\!\!
  d\xi\ {\rm e}^{\Lambda} G(\xi).
\end{equation}
Here we used (\ref{eq:13}) (steady state of (\ref{eq:14})), i.e.
\begin{equation}
  \label{eq:C8}
  \int_{-\infty}^{+\infty}\!\! d\xi\ \xi
  F=\frac{\rho}{2}[\langle s\xi_-\rangle+\langle s\xi_+\rangle]=0.
\end{equation}
Note that using (\ref{eq:C8}) one readily shows that $C_+$ is finite
if (\ref{eq:C4}) holds.  From (\ref{eq:C5}) and (\ref{eq:C6}) it
follows that (\ref{eq:C2}) behaves asymptotically as
\begin{equation}
  \label{eq:C9}
  Q_s(\xi)\sim \left\{
    \begin{array}{ll}
      {\d |\xi|^{-3} \int_{-\infty}^\xi d\xi' \xi' F(\xi') 
        \quad}&\hbox{as  } \xi\to-\infty,\\
      {\d C_+ \xi{\rm e}^{-\Lambda}}
        &\hbox{as  }\xi\to+\infty.
    \end{array}\right.
\end{equation}

To prove (\ref{eq:C2}) and (\ref{eq:C4}) we first note that the
general solution of (\ref{eq:C1}) is
\begin{equation}
  \label{eq:C10}
  Q(\xi)=Q_s(\xi) +C_1 Q_1(\xi) +C_2 Q_2(\xi),
\end{equation}
where $C_1$ and $C_2$ are constants, $Q_1$ and $Q_2$ are two linearly
independent solutions of the homogeneous equation associated with
(\ref{eq:C1}). Two such solutions are
\begin{equation}
  \label{eq:C11}
  Q_1(\xi)= \xi{\rm e}^{-\Lambda},
\end{equation} 
\begin{equation}
   \label{eq:C12}
   Q_2(\xi)= 1- \frac{\xi{\rm e}^{-\Lambda}}{B_1}
   \int_{-\infty}^\xi \!\! d\xi'
   \ \xi'{\rm e}^{\Lambda'}.
\end{equation}
For $\xi<0$, after integration by part $Q_2$ can be expressed as
\begin{equation}
  \label{eq:C13}
  Q_2(\xi)=-\xi{\rm e}^{-\Lambda}
  \int_{-\infty}^\xi \!\! d\xi'
  \ \frac{{\rm e}^{\Lambda'}}{\xi'^2}.
\end{equation}
We now show that the realizability constraint requires that
$C_1=C_2=0$.  First, one readily check that $Q_1$ grows unbounded as
$\xi\to-\infty$, while $\lim_{\xi\to-\infty} Q_2= \lim_{\xi\to-\infty}
Q_3=0$. Hence in order that $Q$ be integrable we must set $C_1$=0 and
the general solution of (\ref{eq:C1}) is
\begin{equation}
  \label{eq:C14}
  Q(\xi)=C_2 Q_2(\xi)+Q_s(\xi).
\end{equation}
For large negative $\xi$, this leads to the expansion
\begin{equation}
  \label{eq:C15}
  Q(\xi)\sim C_2B_1|\xi|^{-3} 
  + |\xi|^{-3} \int_{-\infty}^\xi d\xi'\  \xi' F(\xi').
\end{equation}
For large positive $\xi$, we must distinguish two cases.  If
(\ref{eq:C4}) does not hold, then (using (\ref{eq:C8}))
\begin{equation}
  \label{eq:C16}
  Q(\xi)\sim -C_2B_1\xi^{-3} 
  +\xi^{-3} \int_\xi^{+\infty} d\xi'\ \xi' F(\xi').
\end{equation}
In contrast, if (\ref{eq:C4}) holds, then
\begin{equation}
  \label{eq:C17}
  Q(\xi)\sim -C_2B_1\xi^{-3} 
  + C_+\xi{\rm e}^{-\Lambda}.
\end{equation}
In (\ref{eq:C15})-(\ref{eq:C17}), if non-zero the $C_2$ term at the
right-hand side will dominate the second term. However, since the
$C_2$ term has opposite sign as $\xi \to \pm \infty$, it must be zero,
i.e., we must set $C_2=0$.  This proves that $Q=Q_s$. Furthermore,
since the $F$ term at the right-hand side of (\ref{eq:C16}) is
negative (recall that from (\ref{eq:12}), $F\le0$), this solution must
be rejected in order that $Q$ be non-negative.  Thus (\ref{eq:C4})
must hold.

\end{multicols}


\begin{references}


  
\bibitem[a)]{0} Electronic address: weinan@cims.nyu.edu
  
\bibitem[b)]{0} Electronic address: eve2@cims.nyu.edu
  
\bibitem{eva99} W. E and E. Vanden Eijnden, ``Asymptotic theory for
  the probability density functions in Burgers turbulence,''
  chao-dyn/9901006, submitted to Phys. Rev. Lett.
  
\bibitem{kra99} R. H. Kraichnan, ``Note on forced Burgers
  turbulence,'' chao-dyn/9901023.
 
\bibitem{pol95} A. M. Polyakov, ``Turbulence without pressure,'' Phys.
  Rev. E {\bf 52}, 6183--6188 (1995).
 
\bibitem{bol97} S. A. Boldyrev, ``Velocity-difference probability
  density functions for Burgers turbulence,'' Phys. Rev. E {\bf 55},
  6907 (1997).
  
\bibitem{gokr98} T. Gotoh and R. H. Kraichnan, ``Steady-state Burgers
  turbulence with large-scale forcing'', Phys. Fluids {\bf 10}, 2859
  (1998).
  
\bibitem{ekhma97} W. E, K. Khanin, A. Mazel, and Ya. G. Sinai,
  ``Probability distributions functions for the random forced Burgers
  equation'', Phys. Rev. Lett. {\bf 78}, 1904 (1997); ``Invariant
  measures for the random-forced Burgers equation,'' submitted to Ann.
  Math.
  
\bibitem{rem1} This is true for smooth forcing. Notice however that
  for some situations (like, e.g., for piecewise linear forcing, or
  for zero forcing and piecewise linear initial data) shocks may be
  created at finite amplitude. In this case, a straightforward
  generalization of the argument in Appendix~\ref{sec:C} shows that
  $A$ is non-zero in (\ref{eq:13}), hence $Q\sim A|\xi|^{-3}$ as
  $\xi\to-\infty$.
  
\bibitem{pol98} A. M. Polyakov, private communication (1998).
  
\bibitem{eva299} W. E and E. Vanden Eijnden, ``Statistical theory for
  the stochastic Burgers equation in the inviscid limit,'' in
  preparation.
  
\end{references}
\end{document}